\documentclass[final,5p,times]{elsarticle}

\usepackage{lineno}
\modulolinenumbers[5]

\journal{Astroparticle Physics}

\usepackage{bm}
\usepackage{textcomp}
\usepackage{picture}
\usepackage[normalem]{ulem}

\newcommand{\dd}{\mbox{d}} 

\begin{document}

\begin{frontmatter}

\title{Reconstruction of air-shower parameters for large-scale radio detectors using the lateral distribution}

\author[kit]{D.~Kostunin\corref{cor1}}
\cortext[cor1]{Corresponding author}
\ead{dmitriy.kostunin@kit.edu}
\author[isu]{P.A.~Bezyazeekov}
\author[kit]{R.~Hiller}
\author[kit]{F.G.~Schr\"oder}
\author[isu]{V.~Lenok}
\author[isu]{E.~Levinson}

\address[kit]{Institut f\"ur Kernphysik, Karlsruhe Institute of Technology (KIT), Karlsruhe, Germany}
\address[isu]{Institute of Applied Physics, Irkutsk State University (ISU), Irkutsk, Russia}

\begin{abstract}
We investigate features of the lateral distribution function (LDF) of the radio signal emitted by cosmic ray air-showers with primary energies $> 0.1$~EeV and its connection to air-shower parameters such as energy and shower maximum using CoREAS simulations made for the configuration of the Tunka-Rex antenna array.
Taking into account all significant contributions to the total radio emission, such as by the geomagnetic effect, the charge excess, and the atmospheric refraction we parameterize the radio LDF.
This parameterization is two-dimensional and has several free parameters.
The large number of free parameters is not suitable for experiments of sparse arrays operating at low SNR (signal-to-noise ratios).
Thus, exploiting symmetries, we decrease the number of free parameters based on the shower geometry and reduce the LDF to a simple one-dimensional function.
The remaining parameters can be fit with a small number of points, i.e. as few as the signal from three antennas above detection threshold.
Finally, we present a method for the reconstruction of air-shower parameters, in particular, energy and $X_{\mathrm{max}}$ (shower maximum), which can be reached with a theoretical accuracy of better than 15\% and 30~g/cm$^2$, respectively.
\end{abstract}


\end{frontmatter}

\section{Introduction}
The determination of the composition of the primary particles is one of the most interesting and complicated problems of experimental high-energy cosmic ray physics.
Imaging instruments, particularly, fluorescence or \v{C}erenkov detectors, detect cosmic ray air showers with high precision, but their duty cycle is only in the order of 10\%.
On the other hand, detectors with a full duty cycle, such as particle detectors, until now have poor sensitivity to the shower maximum and cannot provide accurate studies of the composition.
A candidate to solve this dilemma is the radio detection of cosmic rays. 
It probably can reach a precision comparable with air-\v{C}erenkov measurements.
However, it still has a number of important open issues such as efficiency, systematic uncertainties and precision of the energy and shower maximum reconstruction, all also depending on the detector layout.

In the present paper we perform a detailed theoretical study based on a real large-scale detector layout.
We performed about 300 simulations based on the reconstruction of measured high energy Tunka-Rex~\cite{TunkaRex_NIM_2015} and Tunka-133~\cite{Prosin:2014dxa} events.
To simulate air showers we used CoREAS~\cite{Huege:2013vt}, software integrated in CORSIKA, which implements the end-point formalism for calculating radio emission from air showers.
In comparison with previous work made for ideal detectors (see, for example,~\cite{Huege:2008tn,deVries:2010ti}), LOPES~\cite{Apel:2014ika} and LOFAR~\cite{Nelles:2014xaa}, our investigations have several important differences.
First, we reproduce detected events with small uncertainty, thus, our simulation could be compared with signals measured by Tunka-Rex, which, in turn, features an absolute amplitude calibration. 
Second, the geometry of the detector matches modern large-scale setups, i.e. the spacing between antennas is about 200 m.
Finally, we transform the simulated signals applying the real hardware properties of Tunka-Rex (amplifiers, antennas, etc.), and check the sensitivity of selected antennas.
That means, we do a statistical study which gives realistic upper limits for the precision of the reconstruction of air shower properties.
``Upper limits'' because we do not include noise and the precision will be slightly worse when taking into account realistic background (see Appendix).
Therefore, this limit could be reached in the case of large signal-to-noise ratios (SNR). 

The complication in describing the radio LDF originates from the interference of two completely different mechanisms of radio emission: emission due to geomagnetic deflection of charged particles, and the Askaryan (also known as charge-excess) effect.
Adding these two effects causes an asymmetric two-dimensional lateral distribution function (LDF).
There are two obvious approaches to describe this lateral distribution: to use a complex two-dimensional function, or to find some symmetries and rewrite the LDF invariantly.
The first approach was successfully tested in~\cite{Nelles:2014xaa}.
It was shown that the LDF can be described with good accuracy, but the method used in this approach requires a large number of points and, thus, a dense array.
Because of that we used the second, customized approach.
We found a transformation reducing the number of dimensions in the LDF representation to one, converting this function to an azimuthal-symmetric one.
For this we estimate the strength of the asymmetry based on the shower geometry.

In additional to the simulation reproducing real Tunka-133 events, we also performed an ``ideal'' simulation using a symmetrical geometry and a three-dimensional dense detector.
In contrast to the other simulations this is not for a statistical study of air-showers with different $X_{\mathrm{max}}$, but for performing a tomographic study of a mean air-shower (the description of this study is given in Chapter~\ref{chap_asymm_beh}).
In this way, we obtained evidence of a new feature of the radio emission, and found new connections of the parameters of the radio emission with the shower maximum.

\subsection{Geomagnetic coordinate system}
To perform our calculations and later the reconstruction in an invariant way, we will use the so-called geomagnetic coordinate system, a special version of shower coordinates.
The outstanding feature of this system is that the electrical field vector has only two non-zero projections to the axes, the third projection is always close to zero.
The basis of this coordinate system takes the form
\begin{eqnarray}
&& \bm{\hat{\mathrm{e}}}_x = \bm{\hat{\mathrm{V}}} \times \bm{\hat{\mathrm{B}}}\,, \\
&& \bm{\hat{\mathrm{e}}}_y = \bm{\hat{\mathrm{V}}} \times (\bm{\hat{\mathrm{V}}} \times \bm{\hat{\mathrm{B}}})\,, \\
&& \bm{\hat{\mathrm{e}}}_z = \bm{\hat{\mathrm{V}}}\,,
\end{eqnarray}
where $\bm{\mathrm{V}}$ and $\bm{\mathrm{B}}$ are the shower axis and the Earth's magnetic field (hat over vector means normalization, e.g. $\bm{\hat{\mathrm{B}}} = \bm{\mathrm{B}} / |\bm{\mathrm{B}}|$), respectively.
Let us also define useful angles: the geomagnetic angle $\alpha_{\mathrm g} = \angle(\bm{\mathrm{V}},\bm{\mathrm{B}})$ and the geomagnetic azimuth $\phi_{\mathrm g} = \angle(\bm{\hat{\mathrm{e}}}_x,\bm{\hat{\mathrm{e}}}_y)$.

\subsection{Simulation sets}
\label{subsec_simul}
All simulations used in the present paper, have been produced with CoREAS~\cite{Huege:2013vt}.
As the hadronic interaction model we selected QGSJET-II.
As the detector layout we used the setup of the Tunka-Rex experiment, which is located at an altitude of 675~m.
The strength of the geomagnetic field was set to $\approx 60$~{\textmu}T, with inclination and declination of about $72^\circ$ and $-3^\circ$, respectively.
For the incoming direction and energy we used measured Tunka-133 events from 2012/2013.
We selected events satisfying the condition $E_{\mathrm{pr}}\sin\alpha_{\mathrm g} > 0.05$~EeV.
Tunka-133 reconstructs only air-showers with zenith angles $<50^\circ$ due to design restrictions.
That way, as initial parameters we used the energy of the primary particle $E_{\mathrm{pr}}$, the arrival direction $(\theta,\phi)$, and the core coordinates $(x,y)$ on the detector plane.
As the primary particle we used the two possible extreme cases for these energies: protons and iron nuclei.
Due to the high resolution of the Tunka-133 instrument, we can reproduce real events with high accuracy.
The most important unknown parameter in the simulation is the depth of the shower maximum.
Using different random seeds and primary particles (proton and iron) we try to limit the deviation between the shower maximum in the simulations and the measurement by Tunka-133 to less than 30~g/cm$^2$, as this is the precision of Tunka-133~\cite{Prosin:2014dxa}.
For the present work we selected about 300 simulated events of each primary particle, using for each event the simulated shower with the smallest deviation between simulated and real $X_\mathrm{max}$.

Signal transformation and event selection on the detector level are made with the Auger Offline software framework~\cite{Abreu:2011fb}.
We used the pattern of the Tunka-Rex antenna type, which is in first order close to a dipole. 
The frequency range is $30$ -- $80$~MHz.
The event reconstruction pipeline is similar to Tunka-Rex, except for the SNR cuts: we do not add noise to the simulations\footnote{
This decision was motivated in order to avoid the study of the influence of noise on the signal, which is more appropriate in an analysis dealing with real measurements.
Thus, the results we obtain here are a theoretical prediction for large SNR.
For the influence of noise please see Appendix.
}, thus, we put only a threshold on the signal amplitude to reduce the digital noise.

All plots, except Fig.~\ref{prod_peak}, are obtained with the Tunka-Rex layout and Tunka-133 event set.
For Fig.~\ref{prod_peak}, we performed a different simulation, as explained in Section~\ref{chap_asymm_beh}

\section{Asymmetry}
Presently there are a number of mechanisms for air-shower radio emission suggested by theorists~\cite{deVries:2010ti,Scholten:2007ky}.
We will consider only two contributions, which have been proven experimentally and which are the most important and dominant ones: geomagnetically induced transverse currents~\cite{KahnLerche1966} and the Askaryan effect~\cite{Askaryan1962a}.
The complexity of adding these two contributions arises from the different mechanisms of the emission.
While the electrical field of the geomagnetic emission is obtained by integration of charged particles $N_e(h)$ over the height $h$ and lies along the $\mathbf{v}\times\mathbf{B}$ vector, the Askaryan emission is mostly defined by the derivative $N_e'(h) = \dd N_e / \dd h$ and polarized along $\mathbf{v} - \mathbf{V}$, where $\mathbf{v}$ is the velocity of the particle and $\mathbf{V}$ is the shower axis.
In Ref.~\cite{2014PhRvD..90h2003B} good agreement between this simple model and measured data is shown.
In our study we therefore assume that total polarization is a sum of two linear polarized contributions with unknown amplitudes.
This leads to the known azimuthal asymmetry of the lateral distribution of the radio signal~\cite{2014PhRvD..90h2003B,Belletoile:2015rea}.

While in principle the asymmetry can be extracted from polarization measurements for individual events~\cite{Aab:2014esa,Schellart:2014oaa}, in practice, this is difficult when the typical event has just three or for stations above the noise threshold.
Therefore, we developed an approach to approximately correct for the asymmetry based on the event geometry only, since this is measurable with better accuracy for events with only few antenna stations.

\subsection{Origin of asymmetry}
The total electrical field at an antenna at each distance $r$ and azimuth $\phi_\mathrm{g}$ can be represented as vector.
\begin{equation}
\bm{\mathcal{E}}(r) = \bm{\mathcal{E}}_{\mathrm g}(r) + \bm{\mathcal{E}}_{\mathrm{ce}}(r) + \bm{\mathcal{E}}_{\mathrm v}(r)\,,
\end{equation}
where $\bm{\mathcal{E}}_{\mathrm g}$ is a dominantly linearly polarized geomagnetic contribution, $\bm{\mathcal{E}}_{\mathrm{ce}}$ is a radially polarized (like a normal \v{C}erenkov) contribution from the Askaryan effect and $\bm{\mathcal{E}}_{\mathrm v} \approx 0$ is a vertical contribution to the signal.
We neglect the contribution from the vertical component, since the angle between the shower plane and radio wavefront is only $1$--$2^\circ$~\cite{Apel:2014usa}.
As in Refs.~\cite{deVries:2010ti,Scholten:2007ky}, we assume that the amplitude of $\bm{\mathcal{E}}_{\mathrm g}$ and $\bm{\mathcal{E}}_{\mathrm{ce}}$ changes only with distance $r$, but is constant over $\phi_\mathrm{g}$.
For $\bm{\mathcal{E}}_{\mathrm g}$ also, the orientation is constant. Thus,
the signal has the following components in the introduced geomagnetic coordinate system
\begin{eqnarray}
\bm{\mathcal{E}}_{\mathrm g} &=& (\mathcal{E}_{\mathrm g},\,0,\,0) = (\mathcal{E}_0\sin\alpha_{\mathrm g},\, 0\,, 0)\,,\\
\bm{\mathcal{E}}_{\mathrm{ce}} &=& (\mathcal{E}_{\mathrm{ce}}\cos\phi_{\mathrm g},\, \mathcal{E}_{\mathrm{ce}}\sin\phi_{\mathrm g},\, 0)\,,
\end{eqnarray}
where $\mathcal{E}_{\mathrm g} = \mathcal{E}_0\sin\alpha_{\mathrm g} \sim E_{\mathrm{pr}}\sin\alpha_{\mathrm g}$, and $\mathcal{E}_{\mathrm{ce}} \sim E_{\mathrm{pr}}$, and $E_{\mathrm{pr}}$ is the energy of the primary particle.
We assume that the strength of the radio emission depends linearly on the energy of the electromagnetic component (and, consequently, on the total energy) of the air-shower.
The squared amplitude has the form
\begin{eqnarray}
\nonumber
\mathcal{E}^2 &=& (\mathcal{E}_0\sin\alpha_{\mathrm g} + \mathcal{E}_{\mathrm{ce}}\cos\phi_{\mathrm g})^2 + \mathcal{E}_{\mathrm{ce}}^2\sin^2\phi_{\mathrm g}\\
&=& \mathcal{E}_0^2\left( (\sin\alpha_{\mathrm g} + \varepsilon\cos\phi_{\mathrm g})^2 + \varepsilon^2\sin^2\phi_{\mathrm g} \right),
\label{squared_amplitude}
\end{eqnarray}
where the asymmetry is defined as $\varepsilon = \mathcal{E}_{\mathrm{ce}}/\mathcal{E}_0$.\footnote{
In previous work~\cite{Aab:2014esa,Schellart:2014oaa} similar calculations were performed with the notation
\begin{equation}
a \equiv \sin{\alpha_{\mathrm g}}(|\bm{\mathcal{E}}_{\mathrm{ce}}|/|\bm{\mathcal{E}}_{\mathrm g}|) = \varepsilon\,.
\end{equation}
}
As we can see, when $\varepsilon > 0$, the LDF is not azimuthally symmetric
\begin{equation}
\mathcal{E}(r)\biggl|_{\varepsilon > 0} \to \mathcal{E}(r,\phi_{\mathrm g})
\end{equation}
To restore the azimuthal symmetry we define a transformation $\mathsf{\hat K}$ (multiplication factor depending on $r$ and $\phi_\mathrm{g}$) eliminating the azimuthal dependence.
We can obtain the asymmetry $\varepsilon$, if the two components of the electrical field are known
\begin{equation}
\left\{
\begin{array}{l}
\mathcal{E}_x = \mathcal{E}_0\sin\alpha_{\mathrm g} + \mathcal{E}_{\mathrm{ce}}\cos\phi_{\mathrm g}\\
\mathcal{E}_y = \mathcal{E}_{\mathrm{ce}}\sin\phi_{\mathrm g}
\end{array}
\right.\,.
\label{asymm_rec}
\end{equation}
This system can be solved for $\sin\alpha_\mathrm{g} \sin\phi_\mathrm{g} \neq 0$.
\footnote{
From the physical point of view, concerning the asymmetry, we discuss the interference of two components (geomagnetic and Askaryan).
At $\alpha_{\mathrm{g}} = 0$ one of them disappears, consequently, term ``asymmetry'' has no sense anymore. 
At $\sin\phi_{\mathrm g} = 0$ both components become collinear, i.e. $\bm{\mathcal{E}}_\mathrm{g} = k\bm{\mathcal{E}}_{\mathrm{ce}}$, and cannot be distinguished at a single antenna.
}
Solving this system we obtain for the asymmetry $\varepsilon$
\begin{equation}
\varepsilon = \frac{\mathcal{E}_{\mathrm{ce}}}{\mathcal{E}_0} = \frac{\mathcal{E}_y / \sin\phi_{\mathrm g}}{\mathcal{E}_x - \mathcal{E}_y \cot\phi_{\mathrm g}}\sin\alpha_{\mathrm g}
\end{equation}
Similar calculations have been also performed to extract the asymmetry from the experimental data of LOFAR~\cite{Schellart:2014oaa}.
From Eq.~(\ref{squared_amplitude}) it is obvious, that $\mathsf{\hat K}$ takes the form
\begin{eqnarray}
\label{kdef} &&\mathsf{\hat K} = \left(\varepsilon^2 + 2\varepsilon\cos\phi_{\mathrm g}\sin\alpha_{\mathrm g} + \sin^2\alpha_{\mathrm g}\right)^{-\frac{1}{2}}\,,\\
&&\mathsf{\hat K} \mathcal{E}(r,\phi_{\mathrm g}) = \mathcal{E}_{\mathrm{corr}}(r) = \mathcal{E}_0\,.
\end{eqnarray}
Although not used hereafter, as an aside, we want to note that the equation simplifies
in the geomagnetic limit ($\mathcal{E}_{\mathrm g} \gg \mathcal{E}_{\mathrm ce}$),
which was used as an approximation by several experimental analyses normalizing the measured amplitude by $\sin\alpha_\mathrm{g}$
\begin{equation}
\mathsf{\hat K} = \frac{1}{\sin \alpha_{\mathrm g}}\,,\,\, \mathcal{E}_{\mathrm{corr}} = \frac{\mathcal{E}}{\sin \alpha_{\mathrm g}}\,.
\label{kshort}
\end{equation}
Hereafter, we use the full form of $\mathsf{\hat K}$ given in Eq.~(\ref{kdef}), not the approximation of Eq.~(\ref{kshort}).

\subsection{Asymmetry behavior}
\label{chap_asymm_beh}
\begin{figure}[h!]
\begin{center}
\includegraphics[width=1.0\linewidth]{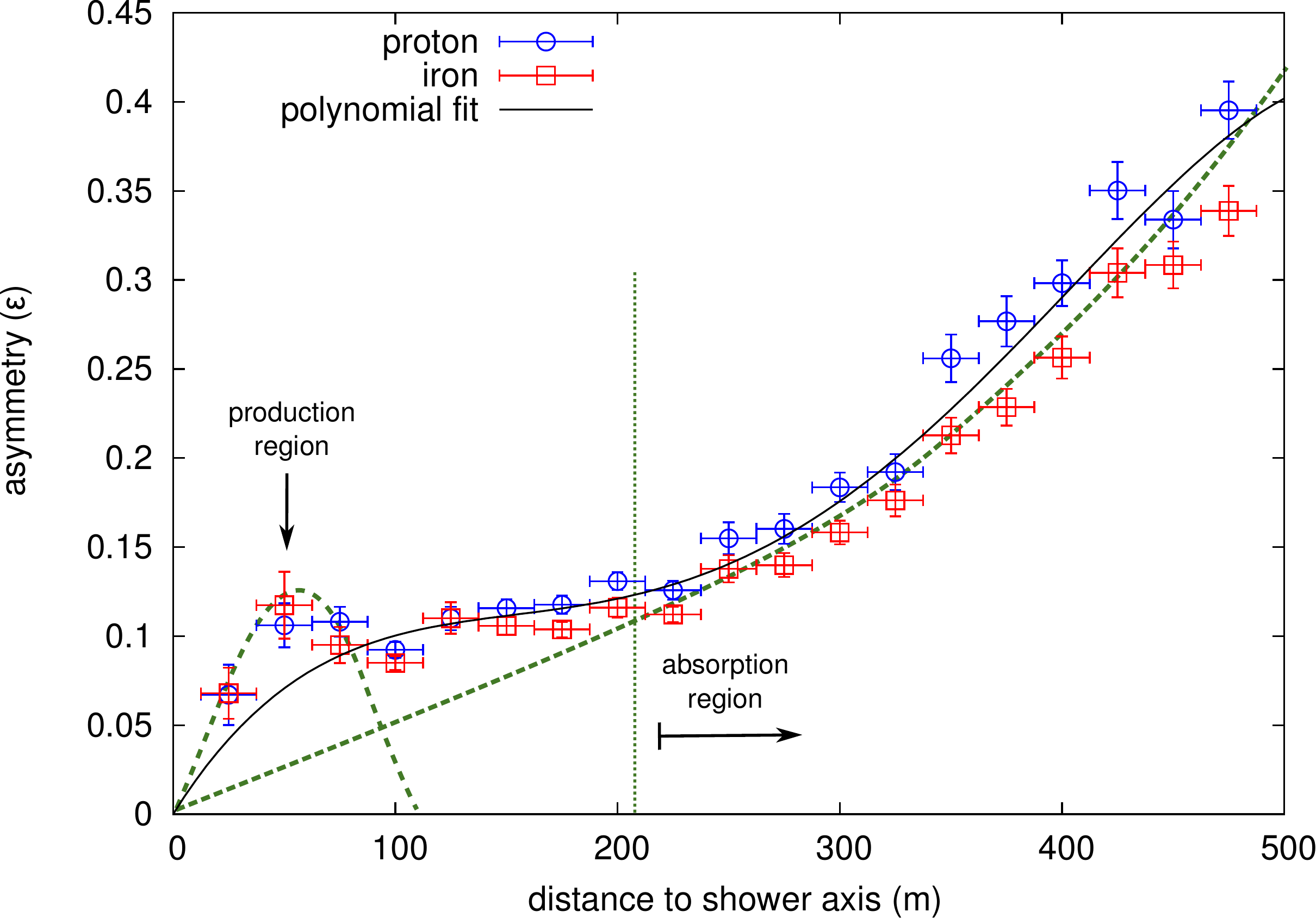}
\end{center}
\caption{Averaged asymmetry profile of the radio lateral distribution for showers initiated by protons and iron nuclei 
based on about 300 CoREAS simulations of Tunka-133 events.
Horizontal bars are the bin size, vertical bars are the standard deviation divided by the square root of the number of entries.
}
\label{iron_asymm}
\end{figure}

For the asymmetry reconstruction we use the formulas given in Eq.~(\ref{asymm_rec}) to study the dependence of $\varepsilon(r)$ on the distance to shower axis $r$.
The distribution of the asymmetry values for different showers is broad.
Thus we created a profile distribution for the average asymmetry values $\varepsilon(r)$ (see Fig.~\ref{iron_asymm}).
The results for proton and iron are similar within about $10\%$.

From this picture we can conclude that the strength of the asymmetry varies not only from shower to shower, but also with distance to the shower axis.
Similar results were obtained in Ref.~\cite{deVries:2013dia}.
It was shown there that the asymmetry has a complicated structure, since geomagnetic and charge excess emission have their maxima at different altitudes, which leads to non-trivial addition of these components.
Taking the longitudinal development of the air-shower into account, de Vries states that the behavior in the near region (up to 150~m from the shower axis) is mostly caused by \v{C}erenkov-like effects.
His statement does not explain the peak in the asymmetry appearing in this region.
We give a potential explanation considering the lateral structure of an air-shower.
The explanation could be the existence of two sources for the charge excess contribution: intensive particle production and inelastic scattering close to the shower axis and particle absorption at distances far from the shower axis (the latter generally agrees with de~Vries).
In other words, in this peak we see the radio emission from the charge excess arising from very intensive particle interaction (like in the classical Askaryan effect in dense media).
The behavior of the production and absorption regions could also depend on the distance to the shower maximum.

To test this statement we studied a simulated radio profile of an air-shower in detail: we simulated a vertical air-shower induced by a proton with energy $10^{17}$~eV. 
The geomagnetic field was set to a strength of about $60$~{\textmu}T (similar to the strength at the Tunka valley) with geomagnetic angle $\alpha_{\mathrm g} = 45^\circ$.
The CORSIKA simulated shower has its maximum at an atmospheric depth of 636 g/cm$^2$.
We put several detector planes at different observation levels from 800 to~1000 g/cm$^2$ with steps of 10~g/cm$^2$.
Each layer consists of concentric rings with radii from 20 to 300~m with steps of 10~m. Each ring consists of 36 antennas placed with azimuthal steps of 10$^\circ$.
By this we obtained a tomographic picture of the shower development in the region after $X_\mathrm{max}$.

\begin{figure}[h!]
\begin{center}
\includegraphics[width=1.0\linewidth]{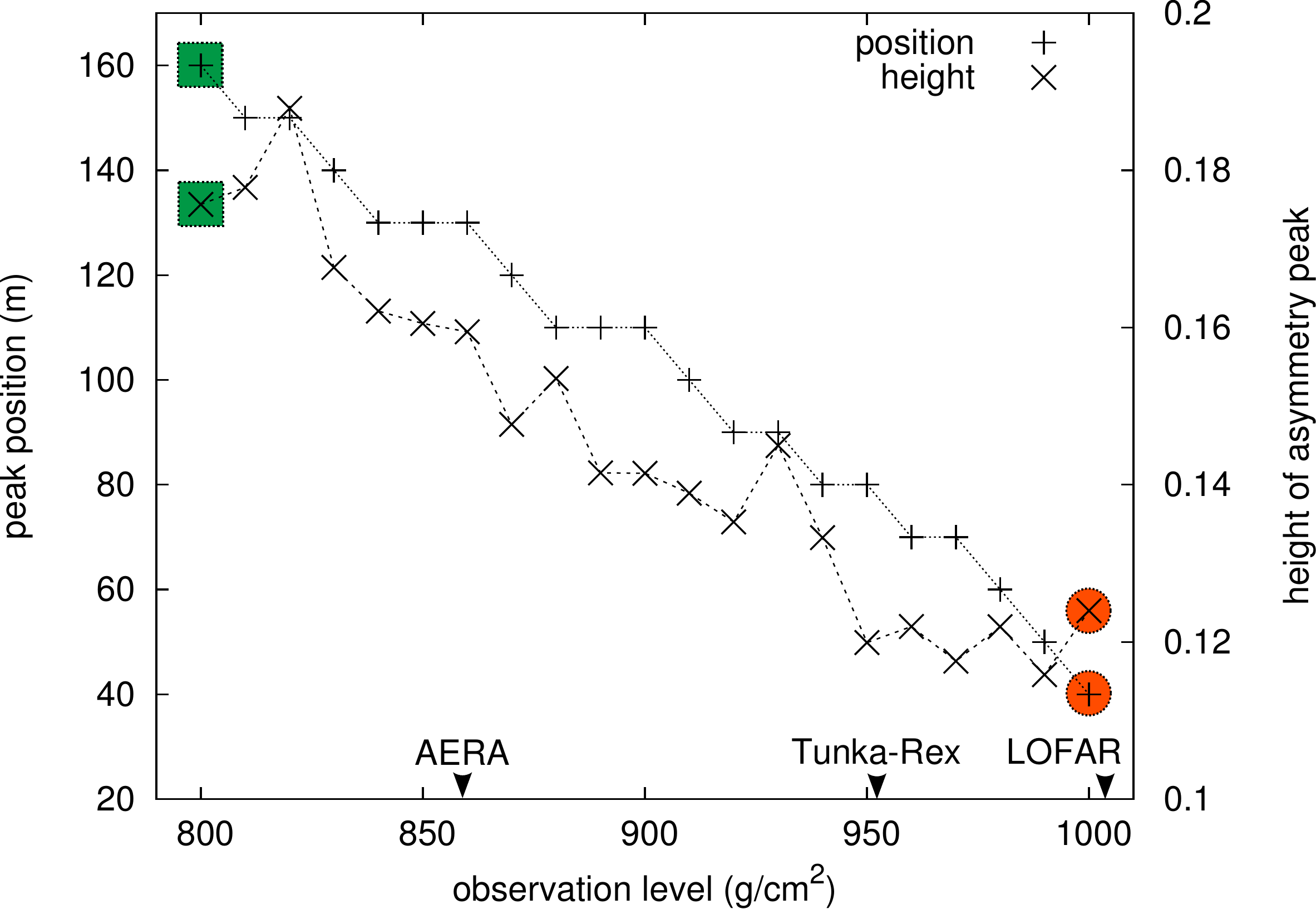}\\
\includegraphics[width=1.0\linewidth]{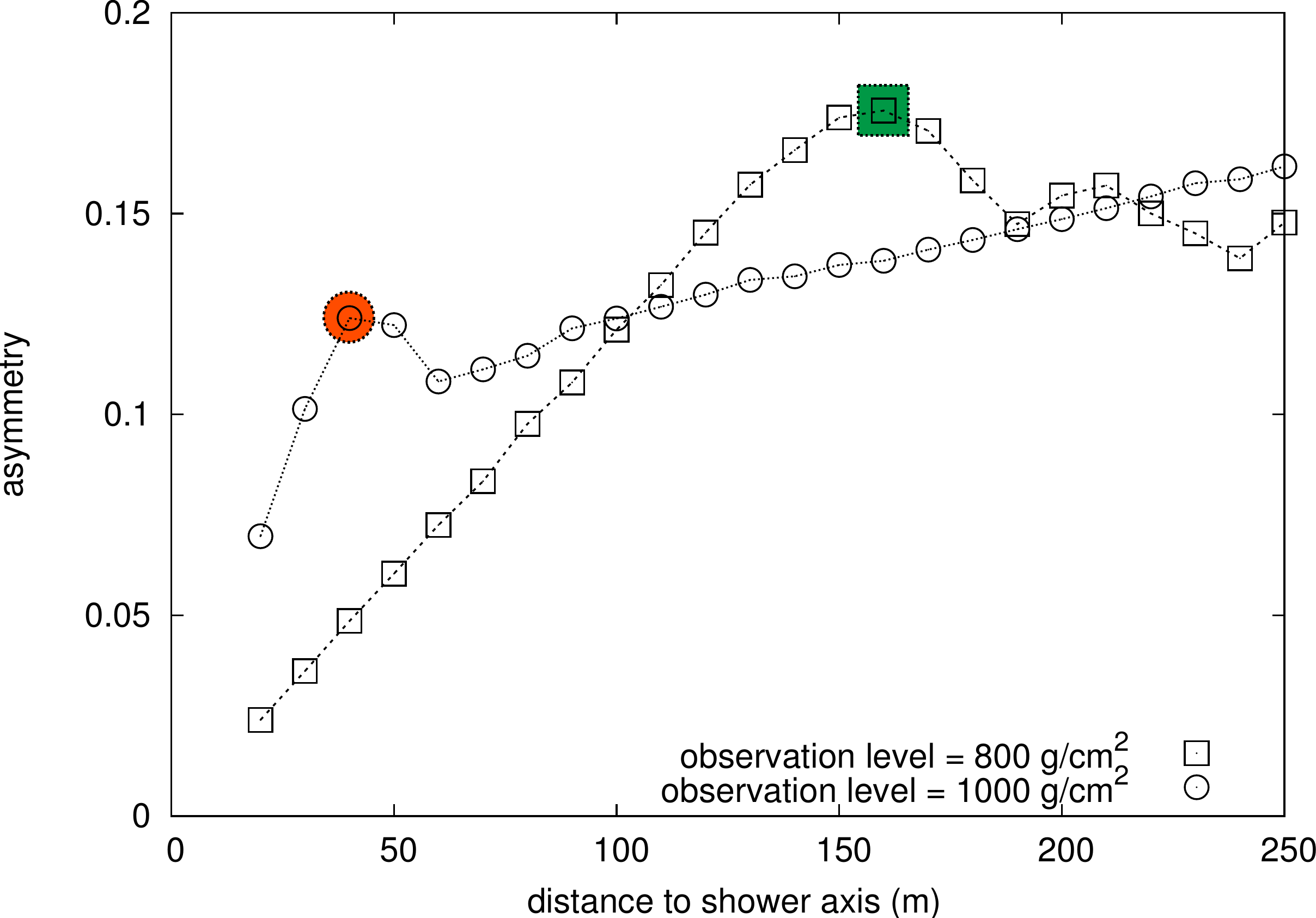}
\end{center}
\caption{\textit{Top:} Correlation between height and position of the asymmetry peak in the production region as function of the observation level of the shower. 
We marked the observation levels for several modern experiments for vertical showers: AERA, Tunka-Rex and LOFAR.
The results are obtained for one exemplary vertical CoREAS shower with a primary energy of 0.1 EeV.
The lines connect the points to guide the eye.
\textit{Bottom:} The distribution of asymmetry against distance to shower axis for two boundary observation levels: 800~g/cm\textsuperscript{2} and 1000~g/cm\textsuperscript{2}.
Highlighted points indicate peaks and show correspondence between top and bottom plots.}
\label{prod_peak}
\end{figure}

The obtained results show that the behavior of the asymmetry in the absorption region does not depend on the shower maximum.
But the position and height of the peak in the production region has a clear correlation with distance to the shower maximum (see Fig.~\ref{prod_peak}).
As we can see, it has the opposite behavior than expected from the \v{C}erenkov-like explanation given by de~Vries~\cite{deVries:2013dia}.
In the case of a \v{C}erenkov-like nature, the peak should move further away from the shower axis with increasing observation level, but we observe the opposite trend.
Possibly, this indicates that after shower maximum the lateral extension of the shower region relevant for Askaryan emission shrinks closer to the shower axis.

In the next step, we search for a simple parameterization of the asymmetry suitable for the reconstruction of measured air-showers.
Although the production region depends on the shower geometry we neglect it, and approximate the asymmetry with a polynomial function $\varepsilon_p(r)$ with fixed point $\varepsilon_p(0) = 0$.
In the present work we do not pay attention to the relation between the peak in the production region and the distance to the shower maximum.
The peak in the production region depends on the shower geometry. Nevertheless, the relative height of the peak is small, especially for deep observers like Tunka-Rex (see Fig.~\ref{prod_peak}), and, therefore, can be neglected in first order.
The absorption region has negligible dependence on shower geometry, therefore, we approximate the asymmetry as a function $\varepsilon_p(r)$ of distance to the shower axis $r$ with fixed point $\varepsilon_p(0) = 0$
\begin{equation}
\varepsilon_p(r) = \sum\limits_{k>0} a^{\varepsilon}_k r^k\,.
\end{equation}
For simplification we set $a^{\varepsilon}_{k>4} = 0$ and performed a global fit for both types of initial particles.
The fit values are given in Table~\ref{asymm_fit_params}.
\begin{table}[h]
\begin{center}
\begin{tabular}{ll}
\hline \hline
Parameter & Value \\
\hline
 $a^{\varepsilon}_1$ & $(2.00 \pm 0.10) \cdot 10^{-3} \mbox{ m}^{-1}$ \\
 $a^{\varepsilon}_2$ & $(-1.37 \pm 0.12) \cdot 10^{-5} \mbox{ m}^{-2}$ \\
 $a^{\varepsilon}_3$ & $(4.10 \pm 0.46) \cdot 10^{-8} \mbox{ m}^{-3}$ \\
 $a^{\varepsilon}_4$ & $(-3.67 \pm 0.56) \cdot 10^{-11} \mbox{ m}^{-4}$ \\
\hline
\end{tabular}
\end{center}
\caption{Global fit values for the asymmetry profile fit in Fig.~\ref{iron_asymm} of averaged CoREAS simulations.}
\label{asymm_fit_params}
\end{table}
One can find the approximate mean asymmetry value by solving the equation $\varepsilon_p''(r) = 0$\footnote{
The statistical mean depends on the choice of the integration range, which can depend on shower geometry.
To obtain a more stable value we decided to take the point of inflexion.
This point is at distances of 100-200 m from the shower axis, where typically most antennas with signal are located, and where $\varepsilon$ is roughly constant.
}.
The asymmetry obtained in this way at $r \approx 150$ m is $\varepsilon_{\mathrm{mean}} = 0.11\pm 0.02$.
This value is in agreement with previous observations~\cite{Aab:2014esa, Schellart:2014oaa}, where also the dependence of $\varepsilon$ over distance to shower axis and zenith was experimentally studied already~\cite{Schellart:2014oaa}.
It is important to note, that in spite of the geometrical invariance of $\varepsilon$ it stills depend on the strength of the magnetic field $\mathbf{B}$, since $\mathcal{E}_\mathrm{g} = \mathcal{E}_\mathrm{g}(\mathbf{B})$.
That means, that the fraction of charge excess $\varepsilon$ should be roughly antiproportional with $|\mathbf{B}|$. Consequently, the found value of $\varepsilon$ for the situation of Tunka-Rex is slightly smaller than the value found by the AERA and LOFAR experiments~\cite{Aab:2014esa, Schellart:2014oaa}.

\begin{figure}[h]
\begin{center}
\includegraphics[width=1.0\linewidth]{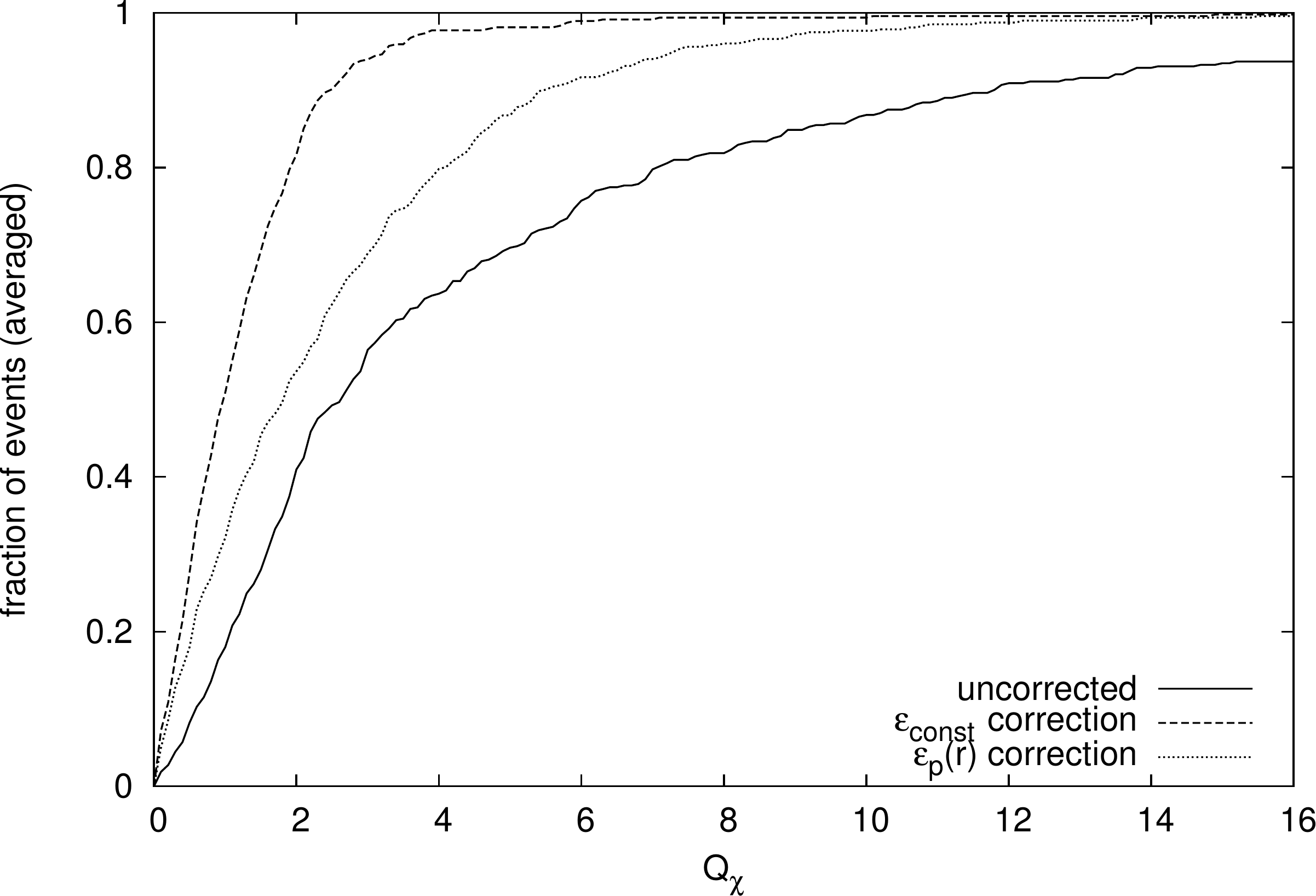}
\end{center}
\caption{Comparison between different methods of correction for about 300 simulated showers.
The fraction of accepted events $\mathsf{N}_\chi(Q_\chi,\varepsilon)$ is calculated for the different forms of LDF $\mathsf{\hat K}(\varepsilon)\mathcal{E}_2(r)$: uncorrected ($\varepsilon$ = 0), constant correction ($\varepsilon = \varepsilon_\mathrm{const} = 0.085$), correction with parameterization ($\varepsilon = \varepsilon(r)$).
Using a constant value for the correction of the azimuthal asymmetry provides the best quality when fitting a one-dimensional LDF.}
\label{correction_quality}
\end{figure}

The next question we studied was: what is the simplest function for description of the asymmetry $\varepsilon$ sufficient for a satisfactory description of the LDF after correction with $\mathsf{\hat K}(\varepsilon)$?
To test the quality of the correction we use a chi-square test.
We define the goodness of the correction by the quantity $\mathsf{N}_\chi(Q_\chi,\varepsilon)$: the fraction of events passing the cut $\chi^2 / \mathrm{NDF} \le Q_\chi$ when the LDF is fitted with $\mathsf{\hat K}(\varepsilon)\mathcal{E}_2(r)$ (with $\mathcal{E}_2(r)$ as defined in next section in Eq.~(\ref{general_1dim_ldf})).
We start with the simplest function, a constant value of the asymmetry $\varepsilon_{\mathrm{const}}$.
To find the optimal value for $\varepsilon_{\mathrm{const}}$ we solve the simple equation
\begin{equation}
\frac{\dd}{\dd \varepsilon}\int \mathsf{N}_\chi(Q_\chi,\varepsilon)\, \dd Q_\chi = 0
\end{equation}
The numerical solution of this equation gives the following values: $\varepsilon_{\mathrm{const}}^{\mathrm{proton}} \approx 0.095$ and $\varepsilon_{\mathrm{const}}^{\mathrm{iron}} \approx 0.075$.
If we compare the goodness of the correction made with this constant asymmetry and the parameterized one (see Fig.~\ref{correction_quality}), we can see that the simple constant function gives a better result.
From the one side, this can be explained by neglecting the peak close to the shower axis in the polynomial parameterization.
The other important factor is large deviations between individual simulated events at distances far from the shower axis (the spread at this distances is indicated by the larger error bars in Fig.~\ref{iron_asymm}), i.e. an effect of shower-to-shower fluctuations.
A quantitative overestimation of the asymmetry can even decrease the goodness of the correction.

The result that a constant value of $\varepsilon$ yields on average an even better correction of the asymmetry than a polynomial makes the practical application of the correction approach very simple.
Consequently, we use a correction by a constant value $\varepsilon = 0.085$ for the further analysis.

\section{Lateral distribution and its connection to the shower parameters}
\label{section_ldf}
\begin{figure}[h!]
\begin{center}
\includegraphics[width=1.0\linewidth]{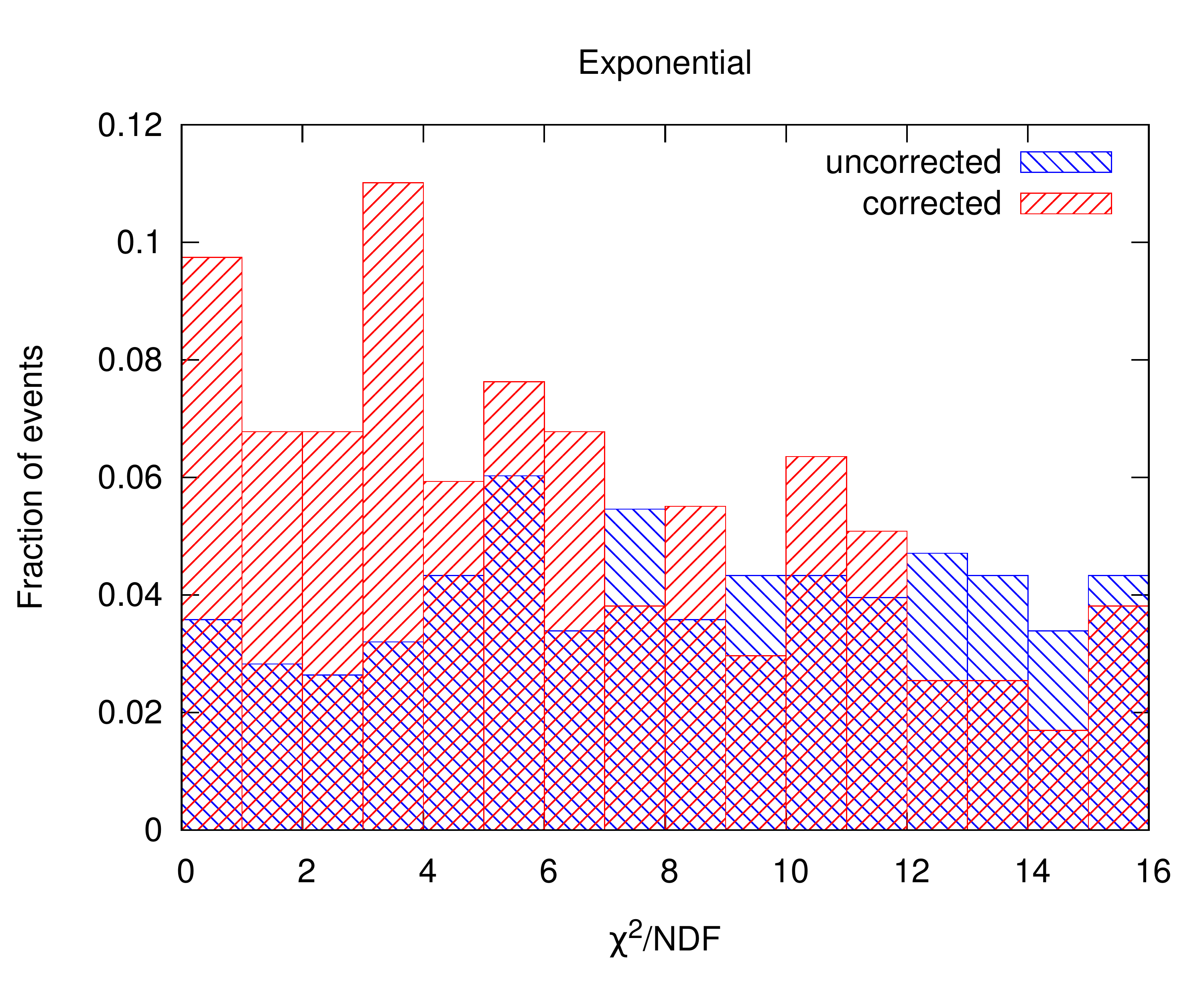}\\
\includegraphics[width=1.0\linewidth]{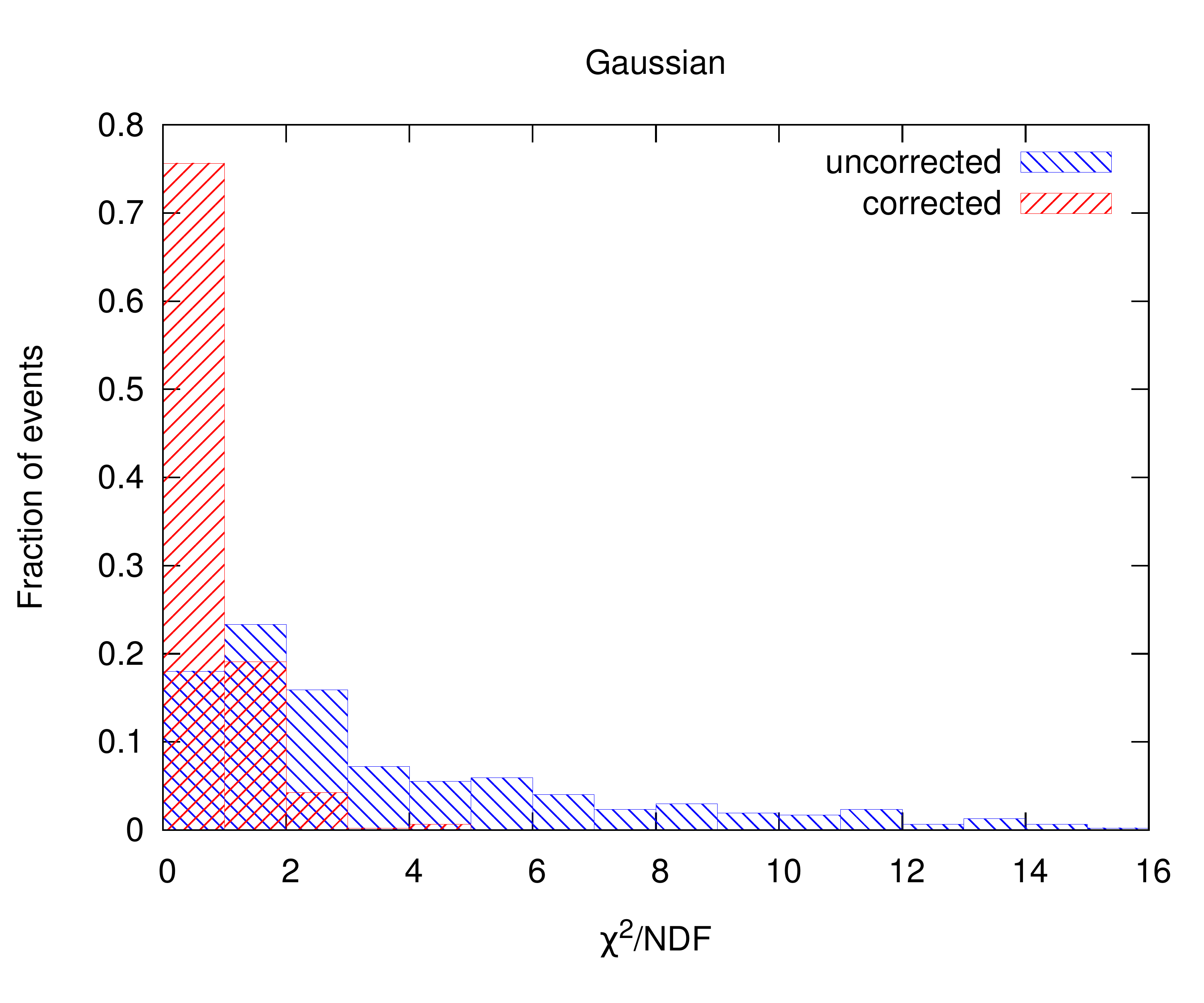}
\end{center}
\caption{Distribution of the quality of LDF fits for different parameterizations (Eq.~\ref{general_1dim_ldf}): exponential ($N=1$) and Gaussian~($N=2$).
The fraction of events with LDFs corrected $\mathsf{N}_\chi(\chi^2/\mathrm{NDF},\varepsilon_\mathrm{const})$ and uncorrected $\mathsf{N}_\chi(\chi^2/\mathrm{NDF},0)$ for the asymmetry is averaged for both primaries (proton and iron).}
\label{corr_diff_par}
\end{figure}

\begin{figure}[h!]
\begin{center}
\includegraphics[width=1.0\linewidth]{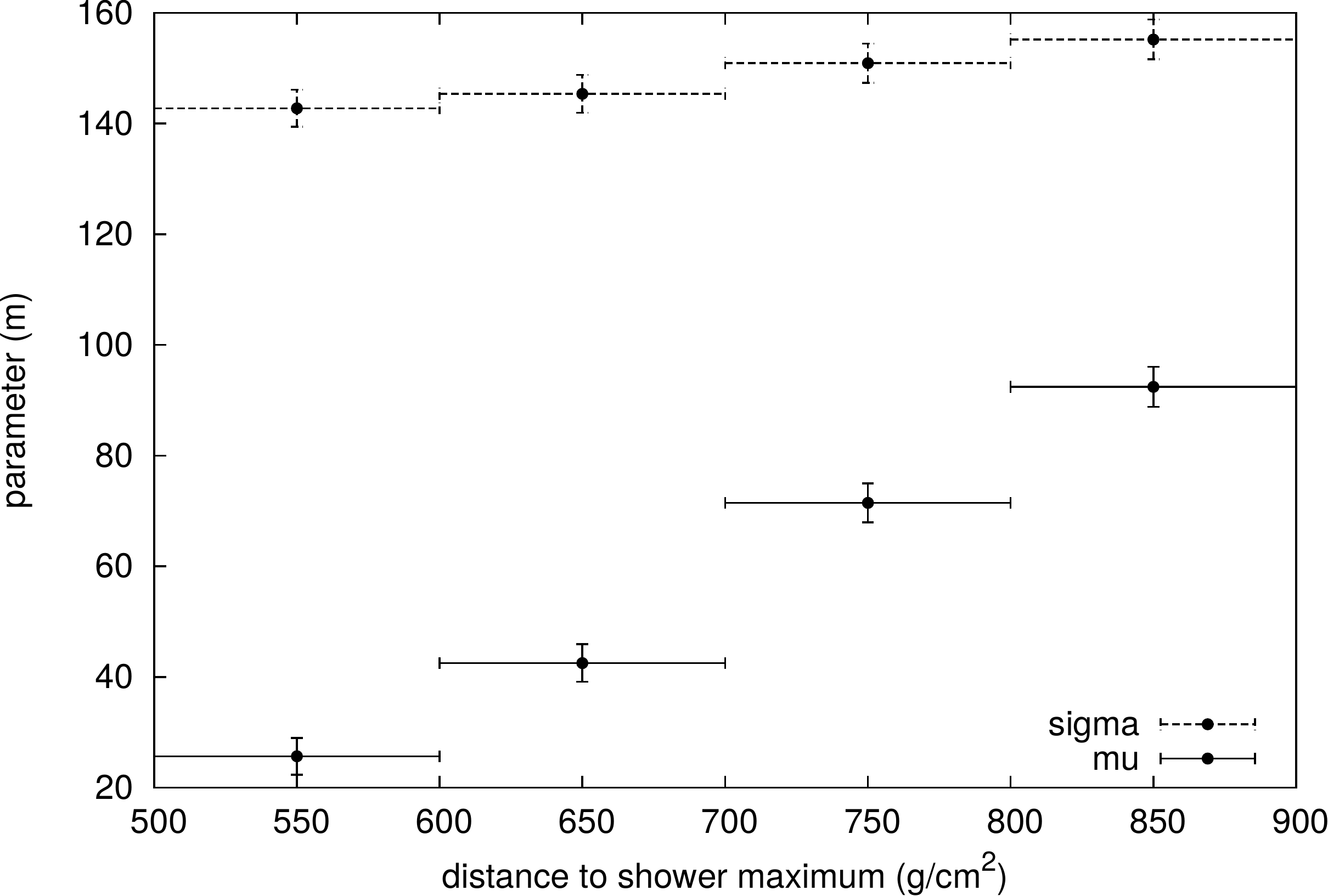}
\end{center}
\caption{Correlation between Gaussian LDF properties and distance to shower maximum. 
Average values for about 300 simulated showers.
Definitions of $\mu$ (mu) and $\sigma$ (sigma) are given in Eqs.~(\ref{sigma_def1}) and~(\ref{sigma_def2}).}
\label{ldf_prop_to_xmax}
\end{figure}

Already in the first observations of the radio emission from air-showers an exponential-like function was suggested as LDF~\cite{1970Natur.227.1116A}.
Later it was confirmed by modern digital experiments, e.g. LOPES~\cite{Apel2010294} and CODALEMA~\cite{Ravel2012S89}.
Thus, we use an exponential-like function written in the form
\begin{eqnarray}
&&\mathcal{E}_N(r) = \mathcal{E}_{r_0} \sin\alpha_{\mathrm g} \exp[f_\eta(r-r_0)]\,,\\
&&f_\eta(x) = \sum\limits_{k=1}^N a_k x^k\,,
\label{general_1dim_ldf}
\end{eqnarray}
with parameters $a_k$ depending on the characteristics of the specific air-shower.
The parameter $r_0$ does not determine the shape of the function, and can be set to a defined, arbitrary value when fitting the LDF (we use $r_0 = r_\mathrm{e} = 120$~m for energy reconstruction and $r_0 = r_\mathrm{e} = 180$~m for $X_\mathrm{max}$ reconstruction).
Using simple considerations (see Ref.~\cite{Scholten:2007ky}, Chapter 2.3.1)  one can state that at the large distances the lateral distribution falls slowly, and the amplitude decreases below any detection threshold.
This means, that the far-distance region is not useful for the reconstruction of the shower maximum.
The slope function $f_\eta$ would have no complicated features in the near-distance region~\cite{Huege:2008tn}, if we did not take into account refraction in the atmosphere (put refractive index $n_r = 1$).
But after introducing the refractive index we immediately obtain \v{C}erenkov-like effects~\cite{Allan:1971,deVries:2013dia}.
For the geometry defined in our simulations, the radius of the \v{C}erenkov ring is at 100-150 m (see Fig.~\ref{ldf_prop_to_xmax}).
This radius can be calculated from Eq.~(\ref{general_1dim_ldf}) solving the equation
\begin{equation}
\frac{\dd}{\dd r} \mathcal{E}_N(r) = 0\,,\,\, r = r(r_0,a_k)
\end{equation}
A solution can be found already with $N = 2$
\begin{equation}
r_{\check{\mathrm{c}}} = r_0 - \frac{a_1}{2a_2}\,.
\end{equation}
We make a comparison between different parameterizations and found that for the selected (Tunka-Rex) geometry a Gaussian (i.e $N=2$) parameterization after asymmetry correction fits almost all events\footnote{
For fitting the LDF we used the true shower axis of the simulations as input.
For Tunka-Rex measurement the shower axis is determined with only small uncertainty by Tunka-133.
}.
For $N=2$, the goodness of fit $\mathsf{N}_\chi(Q_\chi = 1) > 70\%$, while for a simple exponential LDF the goodness $\mathsf{N}_\chi(Q_\chi = 4) \approx 35\%$ (see Fig.~\ref{corr_diff_par}).

The properties of the Gaussian LDF such as mean ($\mu$) and width ($\sigma$) are connected to the distance to the shower maximum (Fig.~\ref{ldf_prop_to_xmax})
\begin{eqnarray}
\label{sigma_def1}
&&\mu = r_{\check{\mathrm{c}}} = r_0 - \frac{a_1}{2a_2}\\
&&\sigma = \frac{1}{\sqrt{-2a_2}}
\label{sigma_def2}
\end{eqnarray}

Now, that we found a good description for the lateral distribution valid for our detector, we tested methods for the reconstruction of air-shower parameters.
We will follow the ideas developed for optical air-\v{C}erenkov emission, because after asymmetry correction the radio emission behaves similarly.

\begin{figure}[t!]
  \setlength{\unitlength}{1.0\linewidth}
  \begin{center}
  \begin{picture}(1.0, 1.0)
  \put(0.57,0.12){\includegraphics[width=0.45\linewidth]{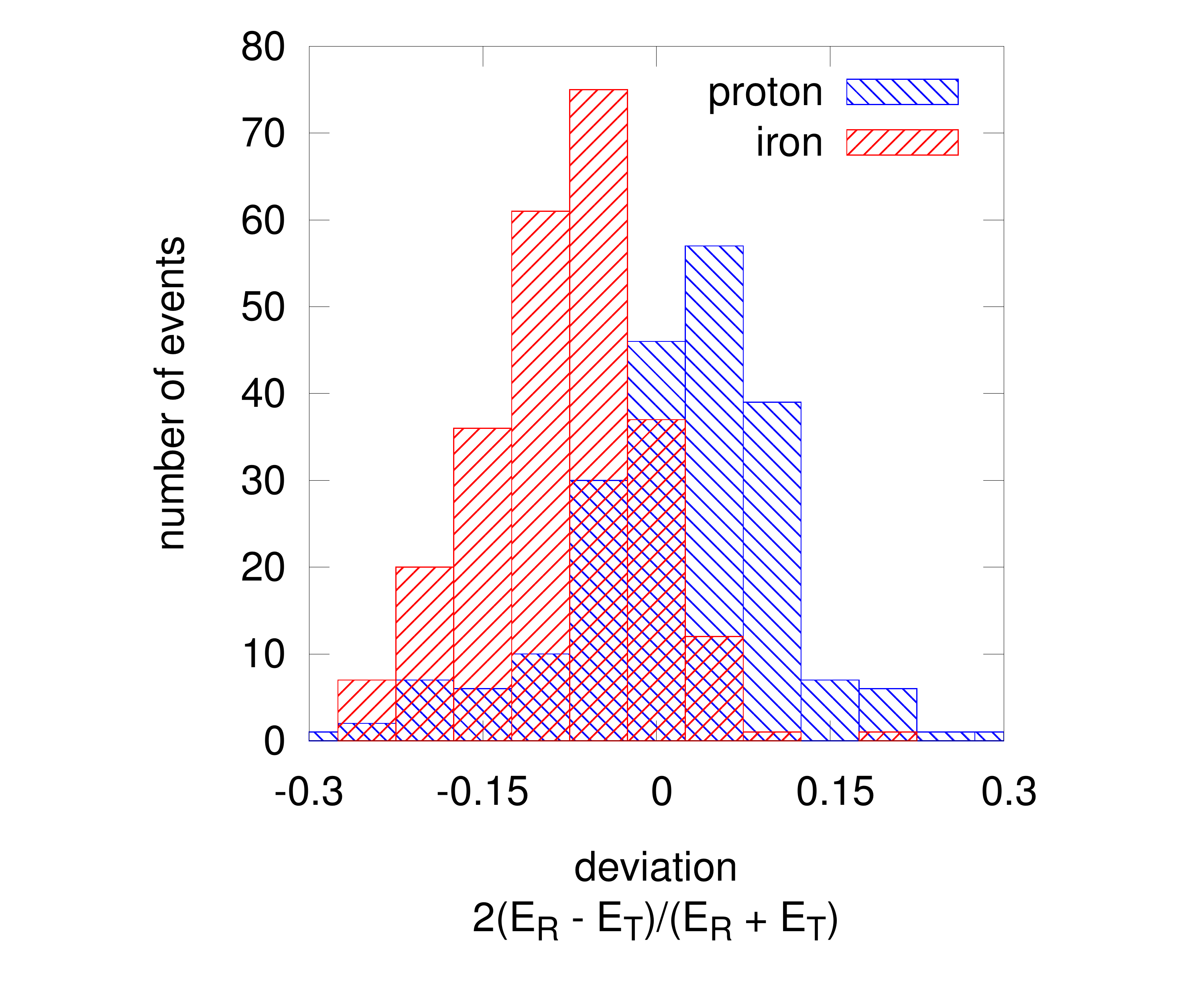}}
  \put(0,0){\includegraphics[width=1.0\linewidth]{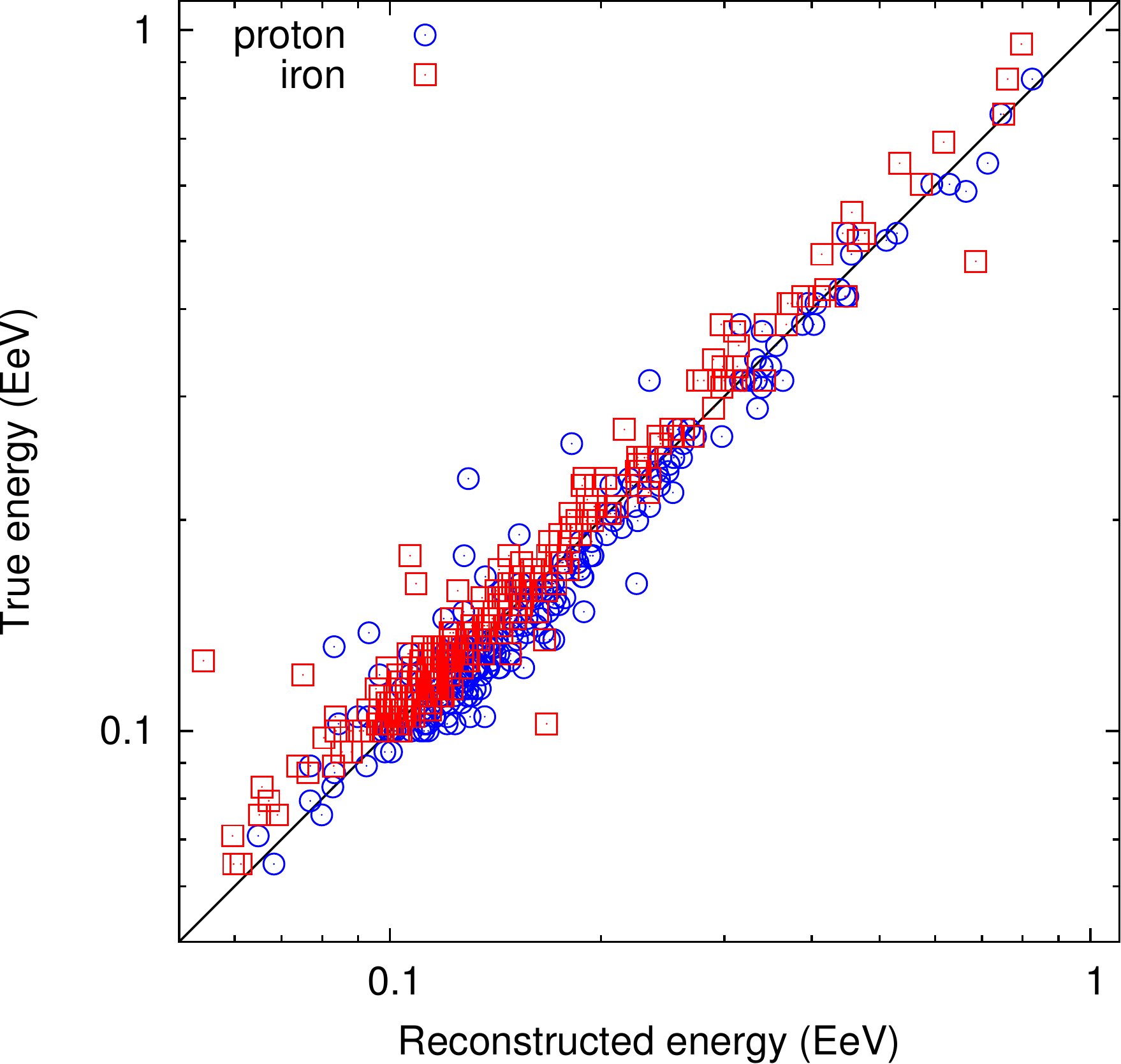}}
  \end{picture}
  \end{center}
  \caption{
  Comparison between true and reconstructed primary energy of the CoREAS simulations.
  }
  \label{reco_comp_amp}
\end{figure}

\begin{figure}[h!]
\begin{center}
\includegraphics[width=1.0\linewidth]{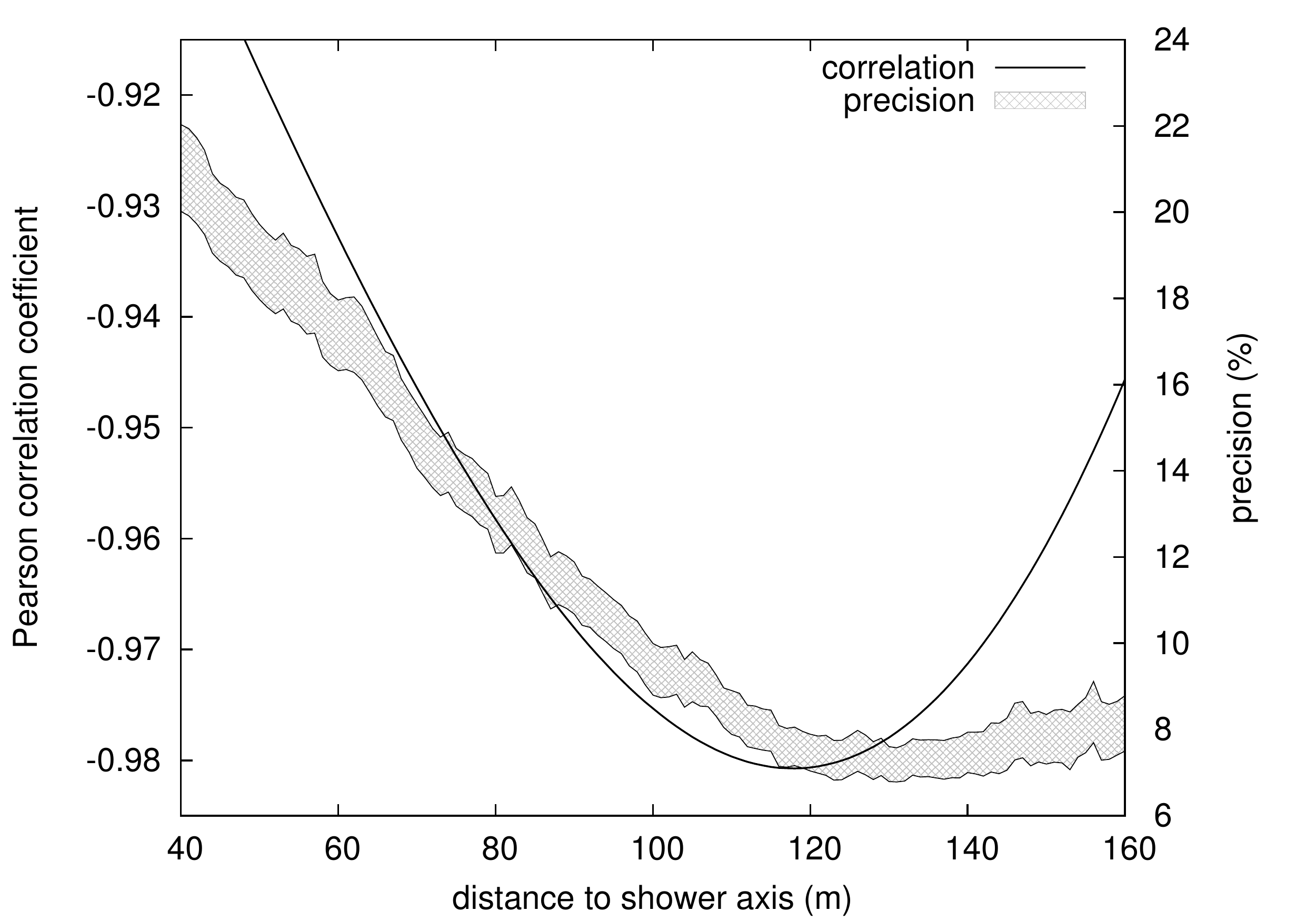}
\caption{Correlation between logarithms of amplitude and energy and precision of the energy reconstruction using Eq.~\ref{energy_rec} at distances from 40 to 160 m.
The curve is the average of proton and iron simulations.
}
\label{corr-prec}
\end{center}
\end{figure}

The energy can be reconstructed by probing the signal amplitude at a defined distance $r_{\mathrm e}$.
Theoretical predictions for the optimal distance are about 50-150 m depending on the mass composition and geometry (see Ref.~\cite{Allan:1971} and later).
Therefore, in general, the energy and LDF are connected by the following phenomenological relation
\begin{equation}
E_{\mathrm{pr}} = \kappa \left(\frac{\mathcal{E}(r_0 = r_\mathrm{e})}{\mbox{V/m}}\right)^b\,,
\label{energy_rec}
\end{equation}
where $\kappa$ is an amplitude slope parameter and $b$ is a power coefficient.
In other words, the energy is proportional to the amplitude at a certain distance $r_\mathrm{e}$.
To simplify experimental data analysis, one can set the arbitrary defined parameter $r_0$ in the fit to $r_\mathrm{e}$ and take the fitted value $\mathcal{E}_{r_0}$ with corresponding fitting uncertainty as energy estimator.
To find $r_\mathrm{e}$ we look simultaneously at the correlation between logarithms of energy and amplitude at different $r_\mathrm{e}$ and at the precision of the energy reconstruction using this formula (see Fig.~\ref{corr-prec}).
The maximum of the correlation points to the distance optimal for the energy reconstruction.
Since the relative difference between optimal distances $r_{\mathrm{e}}^{\mathrm{proton}}$ and $r_{\mathrm{e}}^{\mathrm{iron}}$, and slopes $\kappa^{\mathrm{proton}}$ and $\kappa^{\mathrm{iron}}$ is about 10\% only, we selected median values for the energy reconstruction
\begin{eqnarray}
&& r_\mathrm{e} = 120\mbox{ m}\,,\\
&& \kappa = 422\,\frac{\mbox{EeV}}{\mbox{V/m}}\,,\\
&& b = 0.93\,.
\end{eqnarray}
The precision of the energy reconstruction using these averaged parameters is better than 10\% for both particle types (see Fig.~\ref{reco_comp_amp}).

\begin{figure}[h!]
\begin{center}
\includegraphics[width=1.0\linewidth]{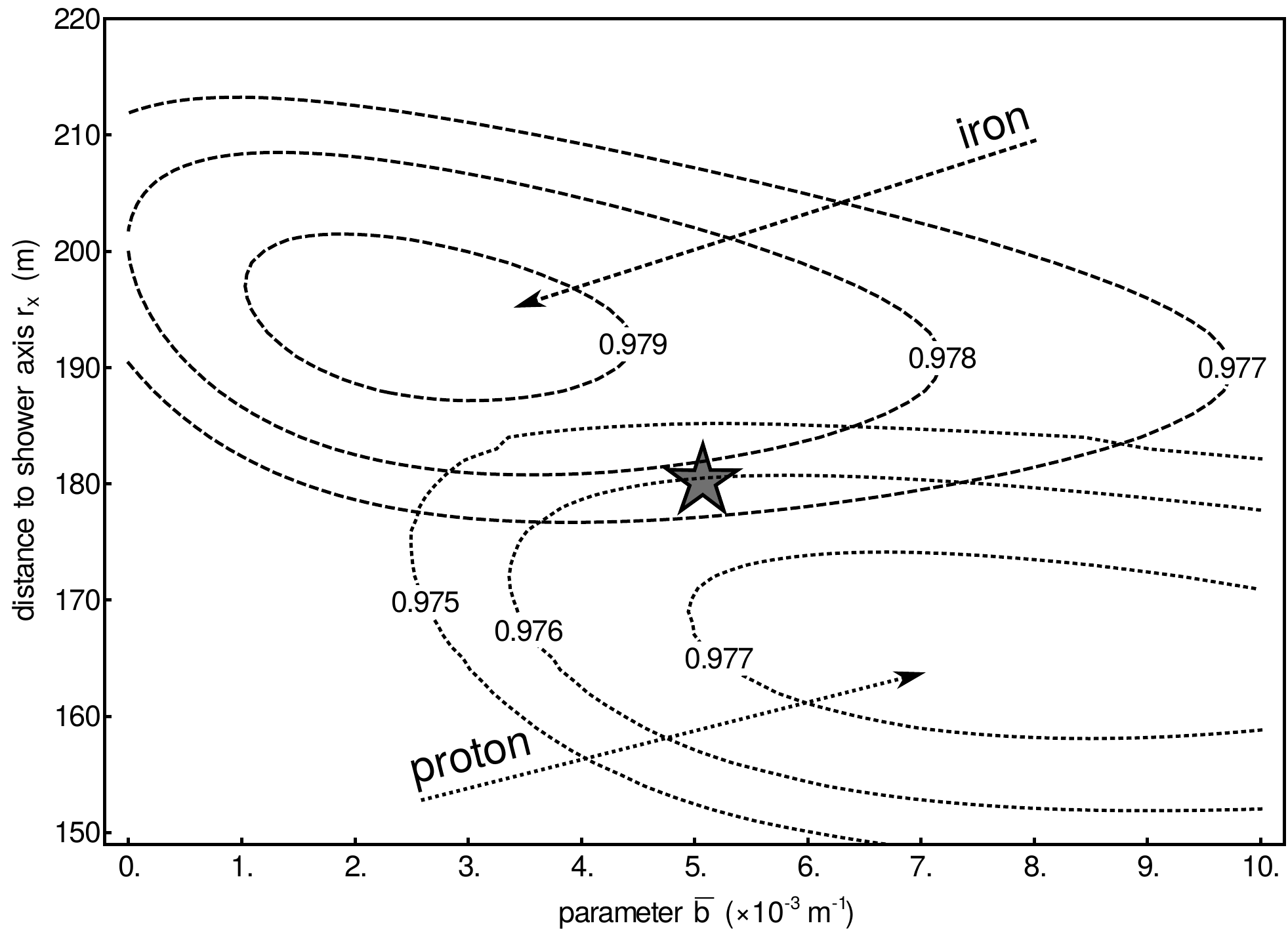}
\end{center}
\caption{
Contour plot of the correlation between LDF slope $\eta$ and true $X_{\mathrm{max}}$ depending on the LDF parameter $r_{\mathrm x}$ and the free parameter $\bar b$.
By fixing $\bar b$, one obtains a distribution similar to Fig.~\ref{corr-prec}. Adding $\bar b$ as free parameter we obtain the two-dimensional dependence on $(\bar b, r_\mathrm{x})$.
In the chosen range of this two-parametric space, the correlation function behaves analytically and converges around $(\bar b = 0.003\mbox{ m}^{-1},\,\,r_\mathrm{x} = 195\mbox{ m})$ and $(\bar b = 0.008\mbox{ m}^{-1},\,\,r_\mathrm{x} = 165\mbox{ m})$ for iron and proton primaries, respectively. 
We have chosen the average point $(\bar b = 0.005\mbox{ m}^{-1},\,\,r_\mathrm{x} = 180\mbox{ m})$ for further analysis, marked as star on the plot.
}
\label{contour}
\end{figure}

\label{xmax_reco}
For the reconstruction of the atmospheric depth of the shower maximum ($X_{\mathrm{max}}$) we use a slope parameter defined as
\begin{equation}
\eta = \frac{\dd f_\eta}{\dd r} = \frac{\mathcal{E}'}{\mathcal{E}} = 2a_2(r - r_0) + a_1\,.
\label{xmax}
\end{equation}
$\eta$ is the slope parameter $a_1$ when evaluating the Gaussian LDF at the distance $r = r_0$
In other worlds, to obtain the slope $\eta$ of the lateral distribution, we take the fitted value $a_1$ after setting $r_0 = r_\mathrm{x}$, since the optimum distance for $X_\mathrm{max}$ reconstruction $r_\mathrm{x}$ is different from the distance $r_\mathrm{e}$ for energy reconstruction.
In Refs.~\cite{Huege:2008tn,deVries:2013dia} a similar method was presented using the slope of the LDF.
For $X_\mathrm{max}$ reconstruction we use the parameterization suggested in Ref.~\cite{Prosin:2014dxa}
\begin{equation}
X_{\mathrm{max}} = X_{\mathrm{det}} / \cos\theta - (A + B\log(a_1 + \bar b))\,.
\end{equation}
This formula is more complicated than the one chosen for energy reconstruction. 
It has two free parameters $A$ and $B$ which will be obtained from a fit to the simulated showers, one distance-dependent parameter $a_1 = \eta(r_{\mathrm x})$, and one correction parameter $\bar{b}$.
We followed the same procedure as for the energy reconstruction: finding the best correlation between the reconstructed and true shower maximum depending on the point $(r_{\mathrm x},\bar b)$ in the two-dimensional space.
One can see the correlation in a contour plot (Fig.~\ref{contour}).

\begin{figure}[h!]
  \setlength{\unitlength}{1.0\linewidth}
  \begin{center}
  \begin{picture}(1.0, 1.0)
  \put(0.55,0.12){\includegraphics[width=0.45\linewidth]{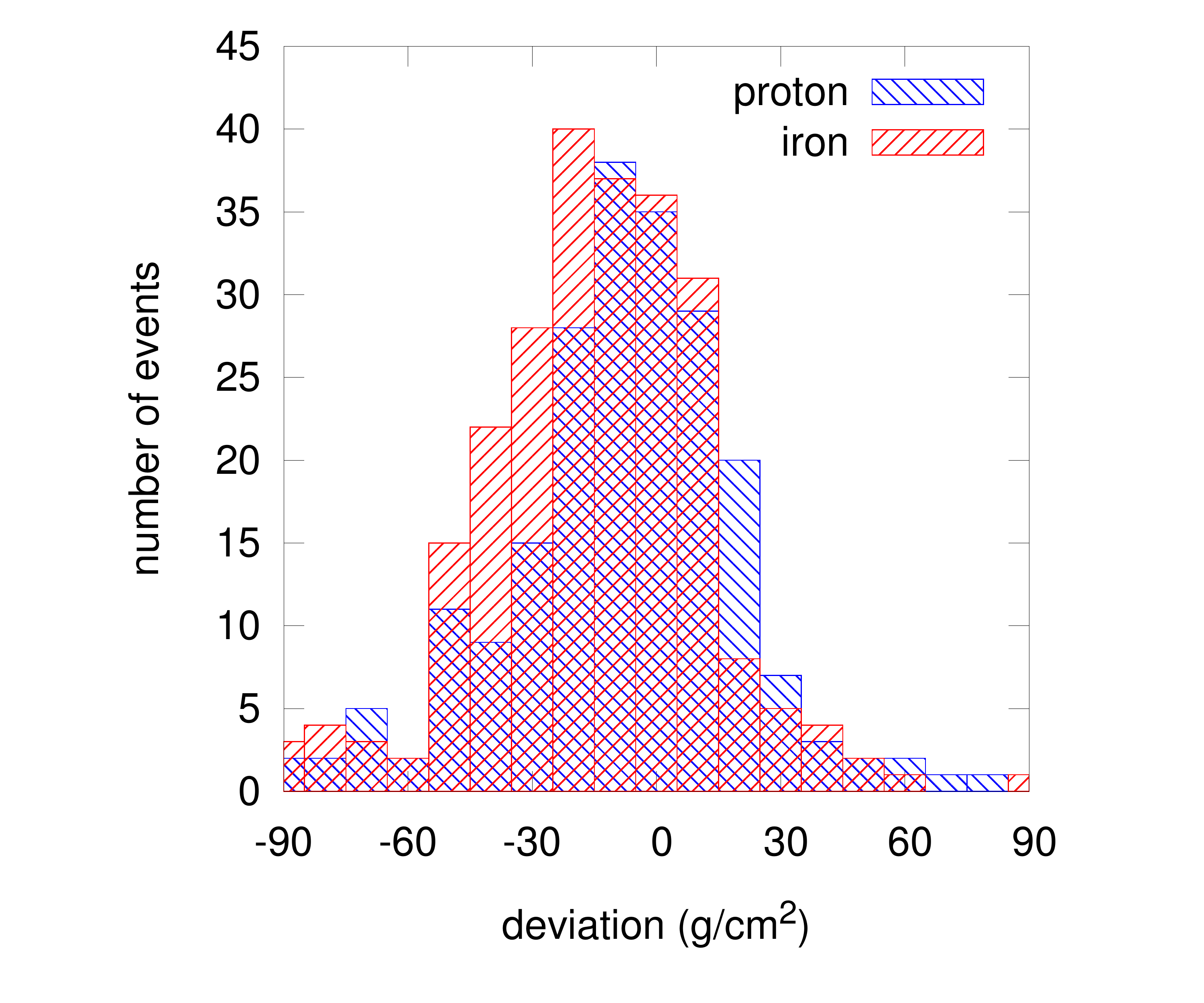}}
  \put(0,0){\includegraphics[width=1.0\linewidth]{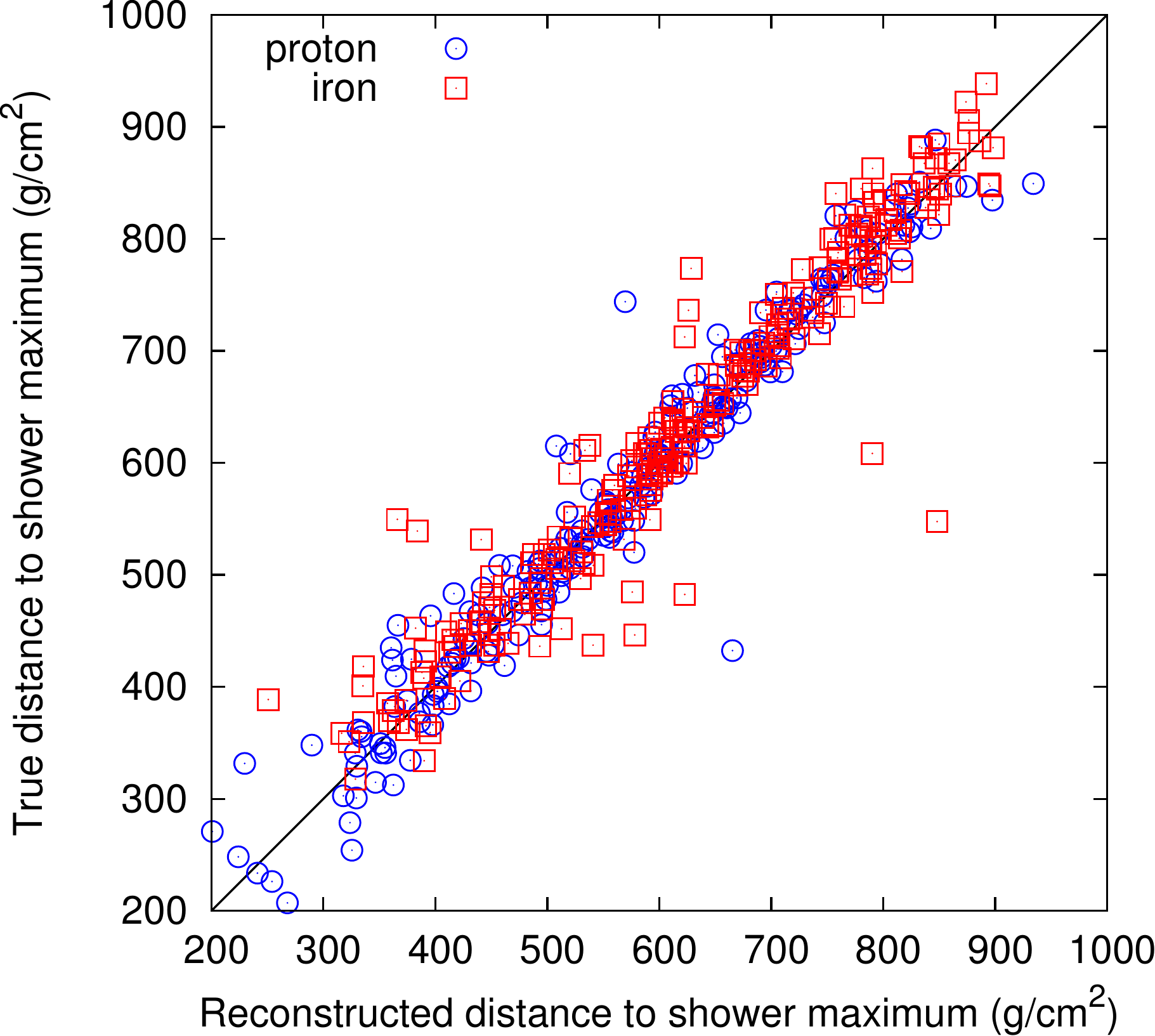}}
  \end{picture}
  \end{center}
  \caption{
  Comparison between true and reconstructed shower maximum of the CoREAS simulations.
  }
  \label{reco_comp_xmax}
\end{figure}

As for the energy reconstruction, parameters in Eq.~\ref{xmax} have about 10\% dependence on the particle type.
After averaging the parameters and applying the formula, the relative difference between true (simulated) and reconstructed $X_{\mathrm{max}}$ values is smaller than 30 g/cm$^2$ (see Fig.~\ref{reco_comp_xmax}).
The averaged parameters are
\begin{eqnarray}
&& r_\mathrm{x} = 180\,\, \mathrm{m}\,,\\
&& A = -1864 \mbox{ g/cm}^2\,,\,\,\, B = -566 \mbox{ g/cm}^2\,,\\
&& \bar{b} = 0.005 \mbox{ m}^{-1}\,.
\end{eqnarray}

\section{Conclusion}
We have found a way to describe the lateral distribution of the radio emission for a large-scale detector with antenna spacing in the order of 200~m.
Our investigation of the azimuthal asymmetry has shown, that, in spite of the complex structure of interference between two different contributions to the radio emission (geomagnetic and Askaryan effects), a simple LDF is sufficient without introducing a large number of arbitrary parameters.
Assuming a constant value for the relative strength of the Askaryan effect provides a sufficient correction for the azimuthal asymmetry of the lateral distribution.
After correction for the asymmetry we need only a one-dimensional LDF with 3 parameters to describe \v{C}erenkov-like effects.
It was shown that for a non-dense radio detector array (with spacing in the order of hundred meters) using a Gaussian LDF is sufficient for primary energies in the EeV range.
For the reconstruction of the energy and shower maximum we used formulas developed for the air-\v{C}erenkov detector Tunka-133.
The free parameters of these formulas were fit to CoREAS simulations made for Tunka-Rex events.
The comparison between the true and reconstructed values of the simulations can be seen in Figs.~\ref{reco_comp_amp} and~\ref{reco_comp_xmax}.

Our results are comparable with the precision of the Tunka-133 host experiment: the precision is better than 15\% for the energy reconstruction and better than 30 g/cm$^2$ for the shower maximum.
The precision can be improved by applying quality cuts.
Let us note that about 20\% of the uncertainty comes from the unknown chemical composition of the primary cosmic rays (i.e. the precisions for a fixed particle types are about 12\% and 25 g/cm$^2$ for energy and shower maximum, respectively).
The reason could be a different distribution of $X_{\mathrm{max}}$ for the proton and iron primaries, or different shower developments.
The developed methods will be used for the data analysis of Tunka-Rex.
The influence of noise has to be studied in more detail since the presented results do not include measurement uncertainties, e.g. due to noise.
At high SNR the theoretically predicted precision should be achievable in practice (see Appendix).

In future, our methods will be further optimized by exploiting information on the polarization, which is connected to the asymmetry.
It is worth noticing that there is no evidence restricting our methods from being applied to more inclined events or events with higher energies, or other detectors with similar antenna spacing and frequency range.

\section*{Acknowledgements}
This work is mainly funded by the German Helmholtz association (grant HRJRG-303) and supported by the Helmholtz Alliance for Astroparticle Physics (HAP).
This work was also supported by the Russian Federation Ministry of Education and Science (G/C 14.B25.31.0010) and the Russian Foundation for Basic Research (Grants 12-02-91323, 13-02-00214, 13-02-12095, 14002-10002).
We are grateful to N.~N.~Kalmykov, V.~V.~Prosin and Andreas Haungs for very fruitful discussions concerning cosmic rays and to Tim Huege for his consultations about CoREAS.

\section*{Appendix: influence of background}
In the present Appendix we show the influence of the noise samples added to simulations on the reconstruction resolution.
As input we used the dataset described in Section~\ref{subsec_simul} and a library of noise measured by Tunka-Rex.
The details of the pipeline used for reconstruction and quality cuts are described in~\cite{TunkaRexXmaxICRC2015}, in particular, we applied a cut on the $\mathrm{SNR}>10$.

The obtained results show, that noise has small influence on energy reconstruction after usual quality cuts (see Ref.~\cite{TunkaRexXmaxICRC2015}).
However, these quality cuts have significant impact on the shower maximum reconstruction.
First, almost all low-energy events are deselected decreasing the statistics of our dataset.
Second, the resolution of the shower maximum reconstruction is slightly worse, about $40$~g/cm\textsuperscript{2}.
This value is in reasonable agreement with measured data~\cite{TunkaRexXmaxICRC2015}.
\begin{figure}[h!]
  \setlength{\unitlength}{1.0\linewidth}
  \begin{center}
  \begin{picture}(1.0, 1.0)
  \put(0.57,0.12){\includegraphics[width=0.45\linewidth]{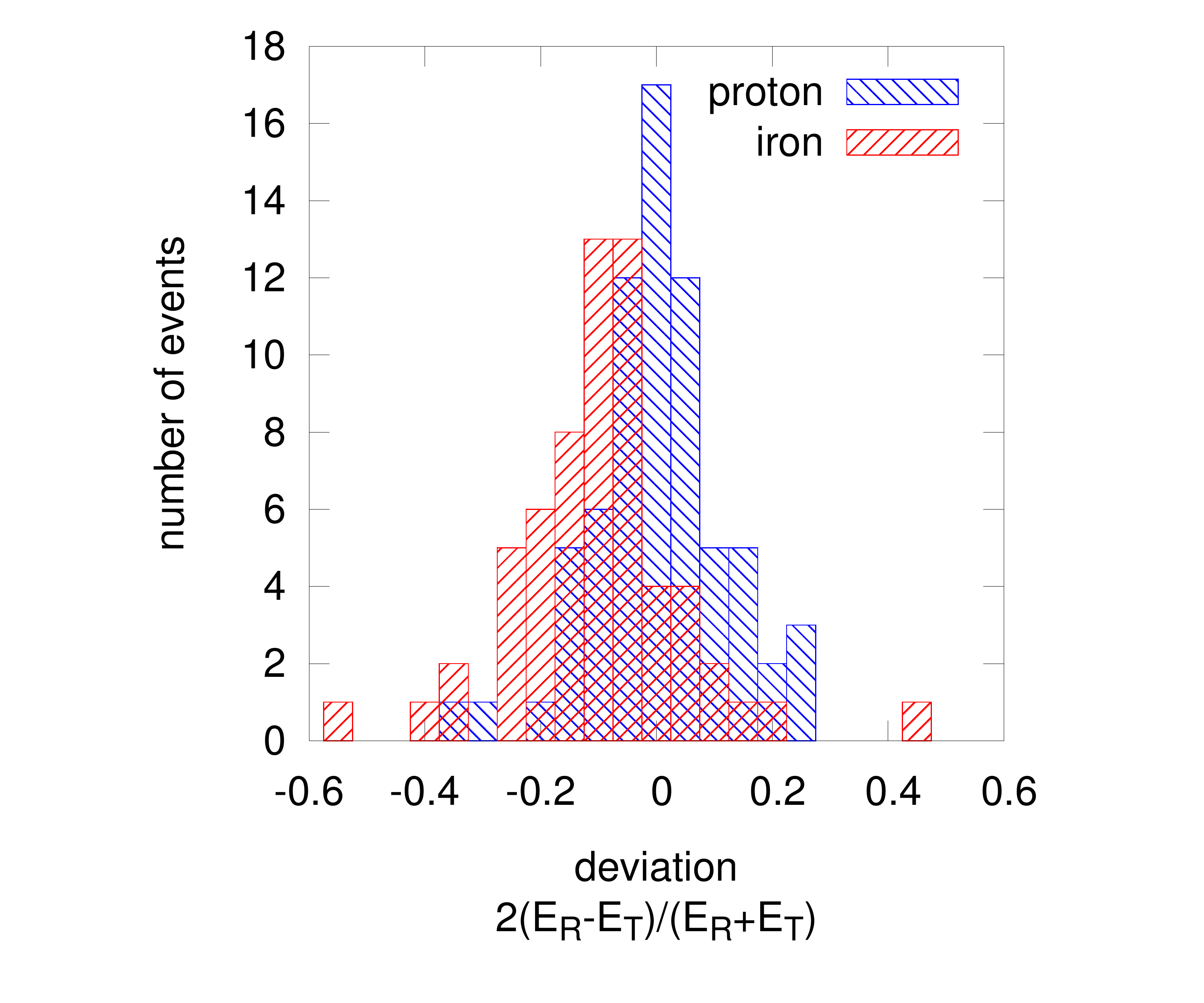}}
  \put(0,0){\includegraphics[width=1.0\linewidth]{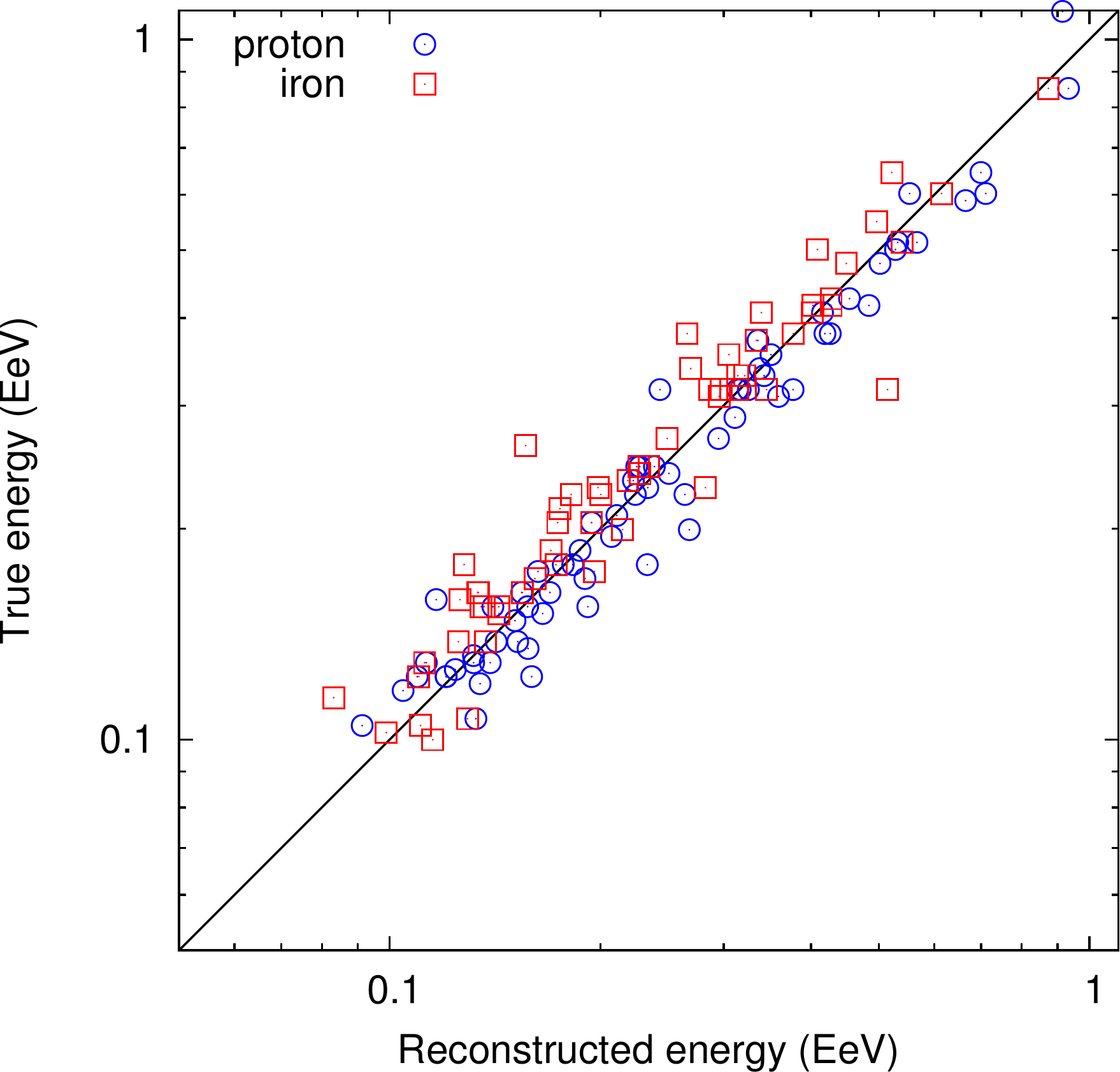}}
  \end{picture}
  \end{center}
  \caption{
  Comparison between true and reconstructed primary energy of the CoREAS simulations including realistic background.
  }
  \label{noise_reco_comp_amp}
\end{figure}

\begin{figure}[h!]
  \setlength{\unitlength}{1.0\linewidth}
  \begin{center}
  \begin{picture}(1.0, 1.0)
  \put(0.55,0.12){\includegraphics[width=0.45\linewidth]{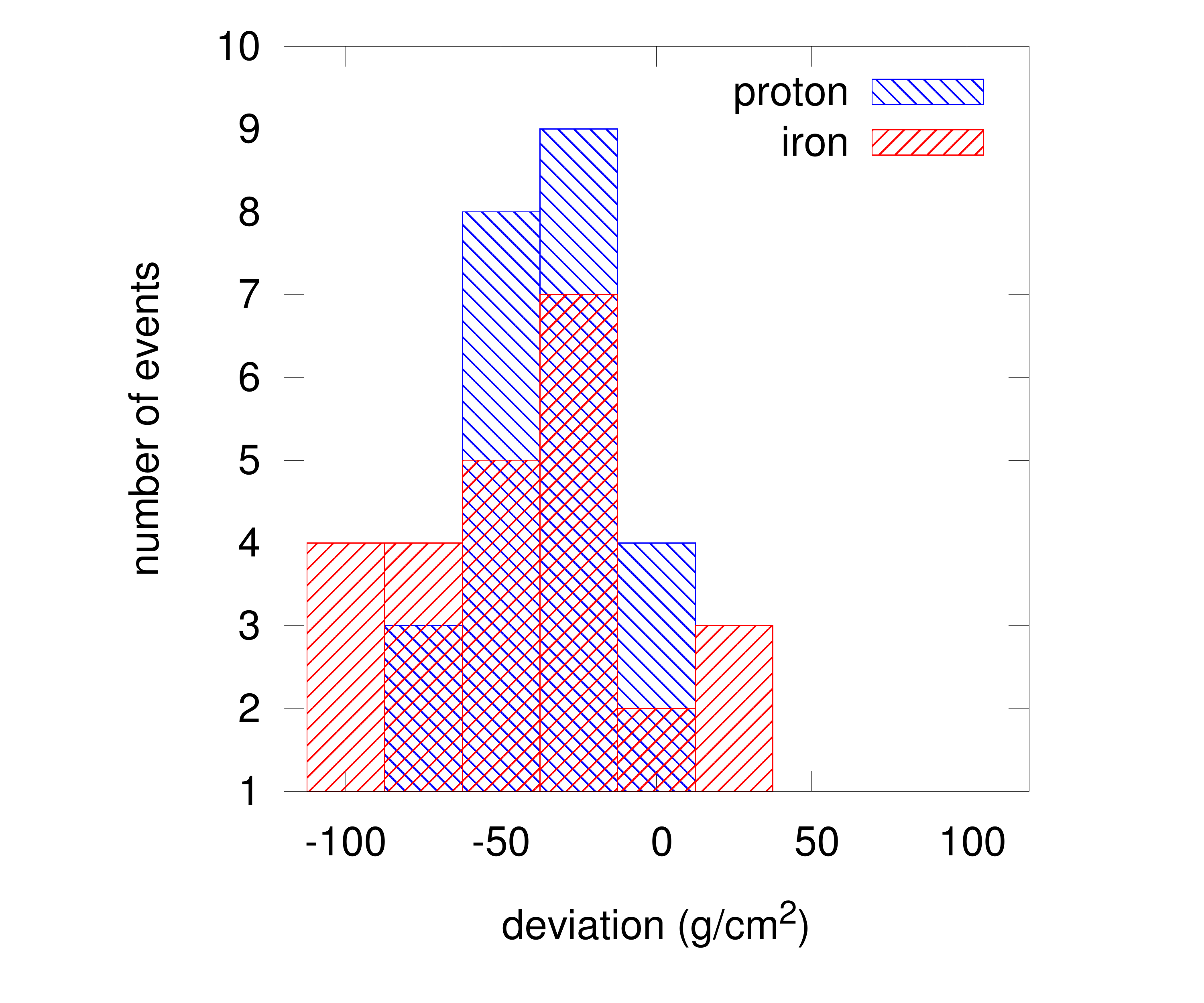}}
  \put(0,0){\includegraphics[width=1.0\linewidth]{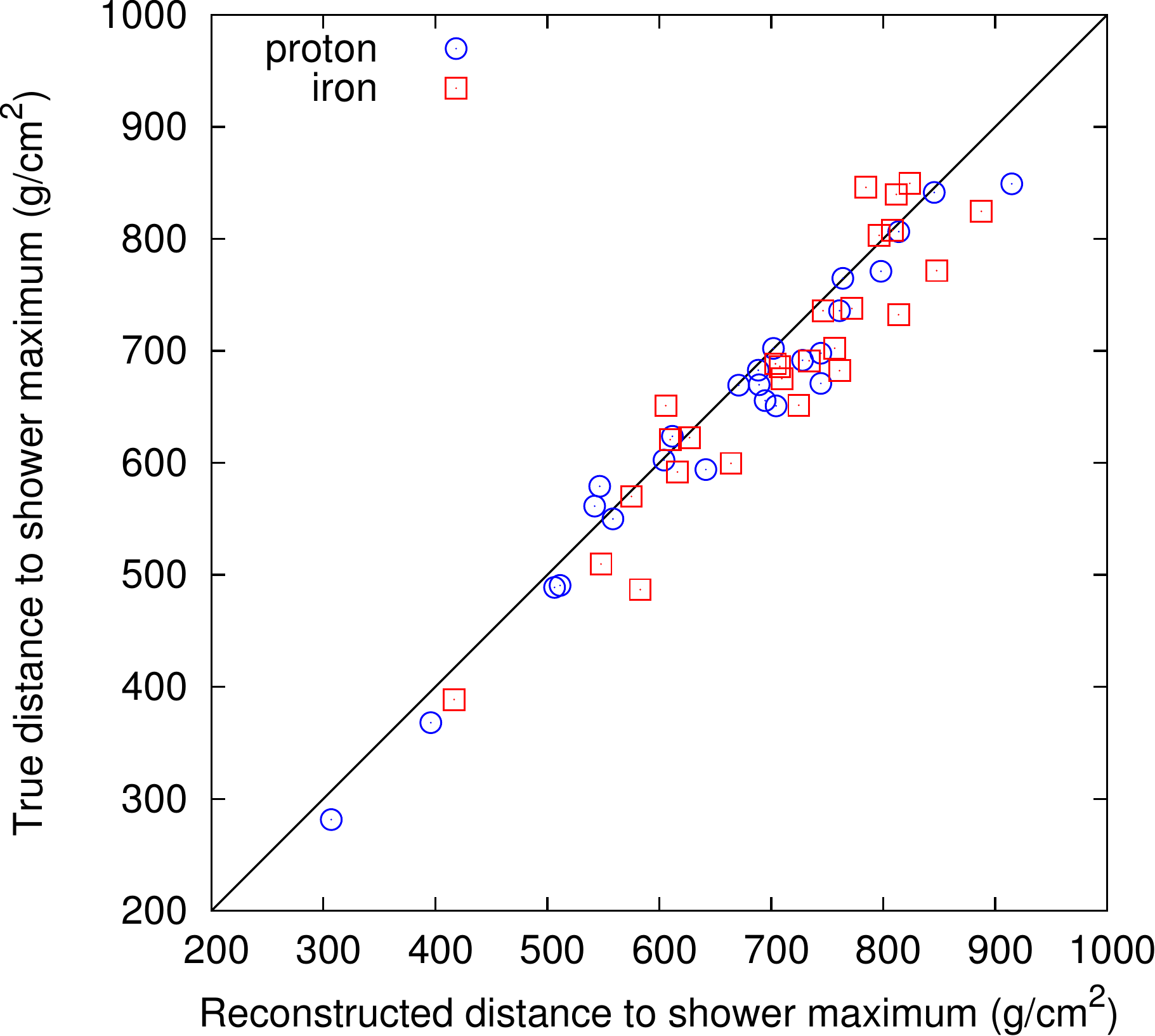}}
  \end{picture}
  \end{center}
  \caption{
  Comparison between true and reconstructed shower maximum of the CoREAS simulations including realistic background.
  }
  \label{noise_reco_comp_xmax}
\end{figure}

\section*{References}

\bibliography{references.bib}

\end{document}